\documentclass[final,5p,times,twocolumn]{elsarticle}
\usepackage{amsfonts}
\usepackage{amsmath}
\usepackage{amsthm}
\usepackage{amssymb}
\usepackage{graphicx}
\usepackage{dcolumn}
\usepackage{bm}
\usepackage{bbding}
\usepackage{mathrsfs}
\usepackage{graphicx}
\usepackage[colorlinks,citecolor=blue,linkcolor=blue,hyperindex,dvipdfm]{hyperref}

\biboptions{numbers,sort&compress}

\begin{document}

\begin{frontmatter}
\title{Effects of the reservoir squeezing on the precision of parameter estimation}
\author{Shao-xiong Wu}
\author{Chang-shui Yu\corref{cor1}}
\ead{quaninformation@sina.com}
\cortext[cor1]{Corresponding author. Tel: +86 41184706201}
\author{He-shan Song\corref{}}
\address{School of Physics and Optoelectronic Technology, Dalian University of
Technology, Dalian 116024, China}
\begin{abstract}
The effects of reservoir squeezing on the precision of parameter
estimation are investigated analytically based on non-perturbation
procedures. The exact analytic quantum Fisher information (QFI) is obtained.
It is shown that the QFI depends on the estimated parameter and its
decay could be reduced by the squeezed reservoir compared with thermal (vacuum)
reservoir, in particular, if the squeezing phase matching is satisfied.
\end{abstract}
\end{frontmatter}

\section{Introduction}

The parameter estimation is one of most important ingredients in various
fields of both the classical and quantum worlds \cite%
{metrology06prl,metrology11np}. The task of quantum estimation is not only
to determine the value of unknown parameters but also to give the precision
of the value. It is a vital issue on how to improve the estimation precision
which is closely related to the quantum Cram\'{e}r-Rao inequality and
quantum Fisher information (QFI) \cite%
{fisher,jiliangxuejiuchan,jiliangxuejiuchan2,wangxiaoguangzongshu,xiaomin,xiangguoyong,kexue,PRA2012}
that determines the bound of the parameter's sensitivity theoretically by
\cite{CR-boud2,Caves94prl}
\begin{eqnarray}
\delta (\phi )\geq1/\sqrt{\nu \mathcal{F}_{\phi }},  \label{CRBound}
\end{eqnarray}%
where $\nu $ means the time of experiments, and
\begin{eqnarray}
\mathcal{F}_{\phi }=\mathrm{Tr}(\rho _{\phi }L_{\phi }^{2})  \label{qfi}
\end{eqnarray}
is the QFI with the symmetric logarithmic derivative $L_{\phi }$ defined by $%
2\partial _{\phi }\rho _{\phi }=L_{\phi }\rho _{\phi }+\rho _{\phi }L_{\phi
} $. Eq. (\ref{CRBound}) implies that the larger QFI means higher
sensitivity of the parameter estimation.

The pioneer work on the quantum parameter estimation were proposed by Caves
\cite{Caves} who showed that the precision of phase estimation can beat the
shot-noise limit (standard quantum limit). Later, lots of jobs with the
similar aims are proposed, such as based on maximally correlated states \cite%
{winelind}, N00N states \cite{tigao-NOON,tigao-NOON2,tigao-coherence},
squeezed states \cite{tigao-squeezed1,tigao-squeezed2}, or generalized
phase-matching condition \cite{xiangweipipei}, and so on. In practical
scenarios, it is inevitable for a quantum system to interact with
environments, the precision of quantum estimation will be influenced by
different extents \cite{Escher,EscherPRL,disange}. In recent years, enormous
effects have been devoted to how to improve the precision of parameter
estimation in the case of open systems. For example, the precision
spectroscopy using entangled state in the presence of Markovian dephasing
\cite{Plenio97} and non-Markovian noise \cite{Plenio12} are investigated;
The QFI under decoherence channels \cite{majian} or in a quantum-critical
environment \cite{sunzhe} are analyzed; The QFI measured experimentally with
photons and atoms are reported \cite{Fishe1,Fishe2}; It is also reported
that the QFI subject to non-Markovian thermal environment \cite{Berrada}
could show revival and retardation loss; The parameter-estimation precision
in noisy systems could be enhanced by dynamical decoupling pulses \cite%
{tigao-wangxiaoguang}, redesigned Ramsey-pulse sequence \cite{tigao-redesign}
or error correction \cite{tigao-errocode} are also shown; Noisy metrology
beyond the standard quantum limits is possible when the noise is
concentrated along some spatial direction \cite{tigao-nosie}. However, if
the environment we considered is a squeezed reservoir, how the QFI
is influenced by the reservoir's parameters?

The squeezed reservoir has been widely studied in quantum information
processing. For example, the squeezed light (reservoir) \cite{Gdadiner} or
finite-bandwidth squeezing \cite{finite-bandwidth,Zoller} for inhibition of
the atomic phase decays and its application in microscopic Fabry-P\'{e}rot
cavity \cite{fb}. In addition, some other considerations of the squeezing
reservoir were also discussed, such as the quantum entanglement dynamics
\cite{guanxisheng}, heat engine recycle \cite{huangxiaoli}, geometric phase
observable \cite{geometricphas}, etc.
The physical realization of the squeezed reservoir has also been proposed
both in theory and in experiment based on various techniques such as the
four-wave mixing \cite{yasuo-fourwave}, the parametric down conversion \cite%
{wulingan}, the suitable feedback of the output signal corresponding to a
quantum nondemolition measurement of an observable \cite{yasuo1}, control
the parameter of the driven laser \cite{yasuo2}, quantum conversion of
squeezed vacuum states \cite{yasuo-canliangzhuanhuan} or energy-level
modulation \cite{yasuo2013}, the atomic systems in cavity QED \cite{yasuo3}
or dissipative optomechanics system \cite{yasuo-yangyaping} and so on. The
reduction of the radiative decay of the atomic coherence in squeezed vacuum
has been realized in the superconducting circuit and microwave-frequency
cavity system \cite{nature}.

In this paper, we will investigate the effects of reservoir squeezing on the
QFI based on the non-perturbation processing \cite{lujingjifen}. We consider
a phase estimation scheme which a two-level system with an imposed unknown
phase interacts with a squeezed reservoir before the final optimal
measurements. To find the influences induced by the reservoir, we derive the
non-perturbative master equation by the path integral method \cite%
{lujingjifen}. In terms of the master equation, we obtain the
exact analytic expression of QFI which is related to the precision of
parameter estimation. It can be found that the QFI depends on the estimated
parameter and the decay of QFI can be reduced by the squeezed reservoir
compared with thermal (vacuum) reservoir. In particular, if the appropriate
squeezing phase matching condition is satisfied, the decay of QFI can be
prevented prominently by the reservoir squeezing.

This paper is organized as follows. In Sec. \ref{sec2}, the parameter
estimation scheme is introduced and the non-perturbation master equation is
obtained. In Sec. \ref{sec3}, the exact analytic expression of QFI for the
estimated parameter is obtained and the effects of reservoir squeezing on
the QFI are investigated. The conclusion are given in the end.

\section{Parameter estimation scheme}

\label{sec2}

\subsection{The scheme}

\begin{figure}[bp]
\centering
\includegraphics[width=1\columnwidth]{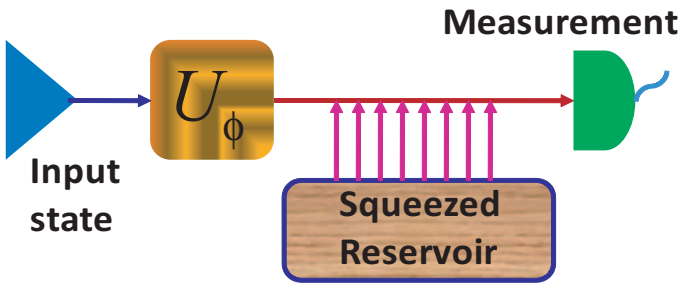}
\caption{ (Color online) The scheme of parameter estimation setup. After
operated by a single qubit phase gate, the system will undergo a squeezed
reservoir. The QFI can be obtained via optimal measurement.}
\label{scheme}
\end{figure}

The setup of the parameter estimation is sketched in Fig. \ref{scheme}. The
input state is a two-level superposed state $\left\vert
\psi_{in}\right\rangle =\left(\left\vert e\right\rangle +\left\vert
g\right\rangle \right)/\sqrt{2} $. After the phase gate ($
U_{\phi}=\left\vert g\right\rangle \left\langle g\right\vert
+e^{i\phi}\left\vert e\right\rangle \left\langle e\right\vert $) is operated on the input
state $\vert\psi_{in}\rangle$, the output state is given by
\begin{eqnarray}
\rho _{out }=U_{\phi}\left\vert \psi _{in}\right\rangle \left\langle \psi
_{in}\right\vert U_{\phi}^{\dagger }.  \label{rhoout}
\end{eqnarray}%
Let the system ($\rho_{out}$) interacts with a squeezed reservoir, the
quantum Fisher information (QFI) of the final state can be obtained via
optimal measurement. The inverse of square root of the QFI is related to the
precision of parameter estimation according to Eq. (\ref{CRBound})
regardless of the experiment times $\nu$.
\subsection{The Hamiltonian}

The initial state of the squeezed reservoir coupled to the system is given
by
\begin{eqnarray}
\rho _{bath} &=&\prod _{k}S_k(r,\theta )\rho _{th}S_k(r,\theta )^{\dagger },
\end{eqnarray}
where the squeezed operator $S_k(r,\theta)$ and the thermal state $\rho_{th}$ are given by
\begin{eqnarray}
S_k(r,\theta ) &=&\exp \left(\frac{1}{2}re^{-i\theta }b_{k}^{2}-\frac{1}{2}%
re^{i\theta }b_{k}^{\dagger 2}\right), \\
\rho _{th} &=&\frac{\exp (-\beta \omega _{k}b_{k}^{\dagger }b_{k})}{\mathrm{%
Tr}\exp (-\beta \omega _{k}b_{k}^{\dagger }b_{k})}.
\end{eqnarray}%
Here $r$ is the squeezed parameter, $\theta$ is the reference phase of
squeezed field and the parameter $\beta =1/(kT)$ with $k$ and $T$ denoting the
Boltzsman constant and temperature, respectively. Noting that the thermal
state $\rho_{th}$ will become the vacuum state $\vert0\rangle\langle0\vert$ if the temperature $T=0$, whilst the environment will become the squeezed vacuum reservoir $\prod_kS_k(r,\theta)\vert0\rangle\langle0\vert S_k(r,\theta)^{\dagger}$ \cite{Breuer,Scully}.

The total Hamiltonian of the system and reservoir reads
\begin{eqnarray}
H={H_{s}}+{H_{bath}}+{H_{int}},
\end{eqnarray}%
with
\begin{eqnarray}
{H_{s}} &=&{\omega _{0}}{\sigma _{z}}/2, \\
{H_{bath}} &=&\sum_{k}{{\omega _{k}}b_{k}^{\dag }{b_{k}}}, \\
{H_{int}} &=&\sum_{k}\left(g_{k}\sigma _{+}b_{k}+g_{k}^{\ast
}\sigma_{-}b_{k}^{\dag }\right),
\end{eqnarray}%
where $\omega _{0}$ denotes the transition frequency of the two-level
system, $\sigma _{\pm }$ is the raising/lowering operators of the system, $%
b_{k}^{\dagger }(b_{k})$ is the creation (annihilation) operators of the
squeezed reservoir and $g_{k}$ is the strength of coupling between the
system and environment.

\subsection{The master equation of reduced density matrix}

In order to get the master equation for the reduced system, we would like to
employ the non-perturbative master equation which can be given, in the Schr%
\"{o}dinger picture, by path integral method \cite{lujingjifen,wangfaqiang}
as
\begin{equation}
\frac{\partial \rho_s }{\partial t}=-iL_{s}\rho_s -\int_{0}^{t}dt^{\prime }{%
\left\langle {{L_{int}}{e^{i{L_{0}}({t}^{\prime }-t)}}{L_{int}}{e^{-i{L_{0}}%
(t^{\prime }-t)}}}\right\rangle _{bath}}{\rho _{s}},  \label{lujingjinfen}
\end{equation}%
where $\rho_s$ denotes the reduced density matrix of the system, $%
\left\langle \cdot \right\rangle _{bath}$ denotes the partial trace of
squeezed reservoir and $L_{0}$, $L_{s}$, $L_{int}$ are the super operators
defined by%
\begin{eqnarray}
{L_{s}}\rho &= &[{H_{s}},\rho ], \\
{L_{0}}\rho &= &[{H_{s}}+{H_{bath}},\rho ], \\
{L_{int}}\rho &= &[{H_{int}},\rho ].
\end{eqnarray}

Assuming the initial state of system plus reservoir is product state $%
\rho_s(0)\otimes\rho_{bath}$, through tedious but straightforward
derivation, the non-perturbative master equation in the interaction picture
can be given by
\begin{eqnarray}
\frac{\partial \rho _{s}}{\partial t} &=&-(N+1)\alpha (t)\left(\sigma
_{+}\sigma _{-}\rho _{s}-\sigma _{-}\rho _{s}\sigma _{+}\right)  \notag \\
&&-(N+1)\alpha ^{\ast }(t)\left(\rho _{s}\sigma _{+}\sigma _{-}-\sigma
_{-}\rho _{s}\sigma _{+}\right)  \notag \\
&&-N\alpha (t)\left(\rho _{s}\sigma _{-}\sigma _{+}-\sigma _{+}\rho
_{s}\sigma _{-}\right)  \notag \\
&&-N\alpha ^{\ast }(t)\left(\sigma _{-}\sigma _{+}\rho _{s}-\sigma _{+}\rho
_{s}\sigma _{-}\right)  \notag \\
&&+2\left(\alpha ^{\ast }(t)M\sigma _{+}\rho _{s}\sigma _{+}+\alpha
(t)M^{\ast }\sigma _{-}\rho _{s}\sigma _{-}\right),  \label{Eq-master}
\end{eqnarray}%
where the coefficients $N$ and $M$ are represented by
\begin{eqnarray}
N &=&n\left(\cosh ^{2}r+\sinh ^{2}r\right)+\sinh ^{2}r, \\
M &=&-\cosh r\sinh re^{i\theta }\left(2n+1\right)
\end{eqnarray}%
with $n=1/(\exp (\beta\omega )-1)$ denoting average photon number.

In this paper, the structure of squeezed reservoir is supposed as the Lorentz form
\begin{equation}
J\left( \omega \right) =\sum_{k}\left\vert g_{k}\right\vert ^{2}\delta
\left(\omega _{0}-\omega _{k}\right)=\frac{\gamma }{2\pi }\frac{\lambda ^{2}%
}{(\omega _{0}-\omega )^{2}+\lambda ^{2}},
\end{equation}%
where $\lambda $ is the spectral width of the reservoir and connects with
the reservoir correlation time as $\tau _{bath}=1/\lambda $, $\gamma $ is
the decay of the system and determines the relaxation time scale as $\tau
_{s}=1/\gamma $. Performing the continuum limit of the bath mode, the time
correlation function can be expressed as \cite{Breuer}
\begin{eqnarray}
\int d\omega J(\omega )e^{i(\omega _{0}-\omega )(t-t^{\prime
})}=\sum_{k}\left\vert g_{k}\right\vert ^{2}e^{-i\omega _{k}(t-t^{\prime })},
\end{eqnarray}
and the coefficient $\alpha(t)$ in the master equation (\ref{Eq-master}) is
\begin{eqnarray}
\alpha (t) &=&\int_{0}^{t}dt^{\prime }\int d\omega J(\omega )e^{i(\omega
_{0}-\omega )(t-t^{\prime })}  \notag \\
&=&\frac{\gamma}{2}\left(1-e^{-\lambda t}\right).
\end{eqnarray}

One can prove that the non-perturbative master equation (\ref{lujingjinfen})
coincides with the second-order time-convolutionless (TCL) master equation
for the two-level system interacting with a thermal reservoir \cite%
{Breuer,lujingjifen}. Because no Markovian approximation is used, it will
lead to non-Markovian dynamics for a qubits coupling to a (squeezed) thermal
reservoir intuitively. However, just like the second-order TCL master equation for
the two-level system interacting with vacuum reservoir, the phenomenon of
temporary backflow of information \cite{hanshimaerkefu,feima-breuer,feima-breuser1,luxiaoming} cannot be revealed.

If the Markovian limit is considered, the characteristic correlation time of
reservoir $\tau _{bath}$ is sufficiently shorter than the system's $\tau
_{s} $, i.e, $\tau _{bath}\ll \tau _{s}$, so the Lorentz spectrum will
become a flat form, i.e., $J(\omega )=\gamma /(2\pi )$, and the coefficient $%
\alpha (t)=\gamma $. Thus, the widely used Markovian master equation \cite%
{Scully,Breuer} can be easily obtained from the Eq. (\ref{Eq-master}),
\begin{eqnarray}
\frac{\partial \rho _{s}}{\partial t} &=&-\gamma (N+1)(\sigma _{+}\sigma
_{-}\rho _{s}-2\sigma _{-}\rho _{s}\sigma _{+}+\rho _{s}\sigma _{+}\sigma
_{-})  \notag \\
&&-\gamma N(\sigma _{-}\sigma _{+}\rho _{s}-2\sigma _{+}\rho _{s}\sigma
_{-}+\rho _{s}\sigma _{-}\sigma _{+})  \notag \\
&&+2\gamma (M\sigma _{+}\rho _{s}\sigma _{+}+M^{\ast }\sigma _{-}\rho
_{s}\sigma _{-}).  \label{markovian}
\end{eqnarray}

\subsection{The solution of master equation}

For the initial system state (\ref{rhoout}), the solution of the master equation given
by (\ref{Eq-master}) can be solved straightforward, which reads
\begin{eqnarray}
\rho _{s}(t)=\left(
\begin{array}{cc}
\rho _{s}^{11}(t) & \rho _{s}^{10}(t) \\
\rho _{s}^{01}(t) & \rho _{s}^{00}(t)%
\end{array}%
\right) ,  \label{rhos}
\end{eqnarray}%
where the elements of density matrix are
\begin{eqnarray}
\rho _{s}^{11}(t) &=&\frac{1}{2}\left( 1+A\right) ,  \notag \\
\rho _{s}^{10}(t) &=&\frac{1}{2}e^{\frac{i\theta }{2}}\left[ \cos \left(\phi
-\frac{\theta }{2}\right)B_{1}+i\sin \left(\phi -\frac{\theta }{2}%
\right)B_{2}\right] ,  \notag \\
\rho _{s}^{01}(t) &=&\rho _{s}^{10}(t)^{\ast },\rho _{s}^{00}(t)=1-\rho
_{s}^{11}(t).  \label{zhongjian}
\end{eqnarray}%
Here, the parameters $A,B_{1},B_{2}$ are given by
\begin{eqnarray}
A &=&\frac{e^{-(1+2n)2\vartheta \cosh (2r)}-1}{(1+2n)\cosh (2r)},  \notag \\
B_{1} &=&e^{-e^{2r}(1+2n)\vartheta },  \notag \\
B_{2} &=&e^{-e^{-2r}(1+2n)\vartheta },  \notag  \label{canshu} \\
\vartheta &=&\int_{0}^{t}d\tau\alpha (\tau ) =\frac{\gamma }{2}\left(t+\frac{%
e^{-\lambda t}-1}{\lambda }\right).
\end{eqnarray}%
It is worth noting that the solution of Markovian master equation (\ref%
{markovian}) can be obtained by just replacing the parameter $\vartheta $ by
$\gamma t$ in Eq. (\ref{canshu}).

\section{Effects of reservoir squeezing on the QFI}

\label{sec3}

\subsection{The analytic QFI}

\begin{figure}[tbp]
\includegraphics[width=1\columnwidth]{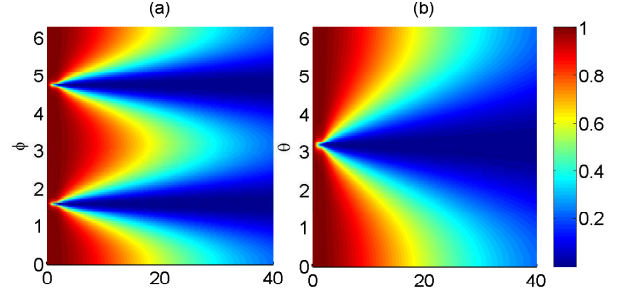}
\caption{(Color online) The dynamics of QFI vs $\protect\gamma t$ and $%
\protect\phi $ ($\protect\theta $). Panel (a) is under the condition $%
\protect\theta =0$, and Panel (b) is under $\protect\phi =0$. For both
panels, $\protect\lambda =0.1\protect\gamma $, $r=1.5$ and $kT=0.5\protect%
\omega $. }
\label{various}
\end{figure}
The explicit expression of the QFI for the estimated parameter $\phi $ is
given by \cite{wangxiaoguangzongshu}
\begin{equation}
\mathcal{F}_{\phi }=\sum_{i}\frac{(\partial _{\phi }\lambda _{i})^{2}}{%
\lambda _{i}}+\sum_{i\neq j}\frac{2(\lambda _{i}-\lambda _{j})^{2}}{\lambda
_{i}+\lambda _{j}}\left\vert \left\langle \varphi _{i}|\partial _{\phi
}\varphi _{j}\right\rangle \right\vert ^{2},  \label{fisher1}
\end{equation}%
where $\lambda _{i}$ are the eigenvalues of estimated state and $\left\vert
\varphi _{i}\right\rangle $ are the corresponding eigenvectors. For pure
states, the QFI can be simplified as $\mathcal{F}_{\phi }=4\left(
\left\langle \partial _{\phi }\varphi |\partial _{\phi }\varphi
\right\rangle -\left\vert \left\langle \partial _{\phi }\varphi |\varphi
\right\rangle \right\vert ^{2}\right) $. Substituting the estimated state (%
\ref{rhos}) into the formula of QFI (\ref{fisher1}), the first term of Eq. (%
\ref{fisher1}) is $\left(B_{1}^{2}-B_{2}^{2}\right)\sin ^{2}\left(2\phi
-\theta \right)/\left[m(2-m)\right]$ with $m=2\left[A^{2}+B_{1}^{2}\cos
^{2}(\phi -\frac{\theta }{2})+B_{2}^{2}\sin ^{2}(\phi -\frac{\theta }{2})%
\right]$ and the second term is $\left\{2A^{2}\left[B_{1}^{2}\sin ^{2}(\phi -%
\frac{\theta }{2})+B_{2}^{2}\cos ^{2}(\phi -\frac{\theta }{2})\right]%
+2B_{1}^{2}B_{2}^{2}\right\}/m$. Summing the two terms, we can obtain he
analytic expression of QFI for the estimated parameter $\phi $
\begin{eqnarray}
\mathcal{F}_{\phi } =\frac{%
B_{1}^{2}(A^{2}+B_{2}^{2}-1)-(1-A^{2})(B_{2}^{2}-B_{1}^{2})\cos ^{2}(\phi -%
\frac{\theta }{2})}{A^{2}+B_{1}^{2}\cos ^{2}(\phi -\frac{\theta }{2}%
)+B_{2}^{2}\sin ^{2}(\phi -\frac{\theta }{2})-1},   \label{fisher}
\end{eqnarray}%
where the parameters $A$, $B_{1}$, $B_{2}$ are given in Eq. (\ref{canshu}).
From the analytic QFI (\ref{fisher}), one can obviously find that the QFI
depends on the estimated parameter $\phi $ and squeezed phasing parameter $%
\theta $. It varies as $\phi $ with the periodicity $\pi $, and as $\theta $
with periodicity $2\pi $ . A vivid illustrations of such relations are given
by Fig. \ref{various}.

\begin{figure}[b]
\centering
\includegraphics[width=1\columnwidth]{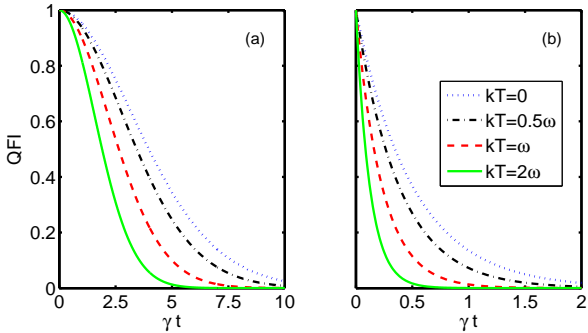}
\caption{(Color online) QFI vs. $\protect\gamma t$ with various temperatures
without squeezing. Panel (a) is based on the master equation (\ref{Eq-master}) with $\protect\lambda =0.1\protect\gamma $ and Panel (b)
respects to the Markovian master equation (\protect\ref{markovian}). Here we
set $kT=0$ (blue dotted line), $0.5\protect\omega $ (black dotted-dashed
line), $\protect\omega $ (red dashed line), 2$\protect\omega $ (green line),
respectively.}
\label{temperature}
\end{figure}

\subsection{The case without squeezing}

In order to show the effects of the reservoir squeezing, we will first give a brief
demonstration of the behavior of QFI without squeezing. That means we
consider the reservoir as a standard thermal reservoir. In this case, the
QFI given in Eq. (\ref{fisher}) can be simplified as
\begin{eqnarray}
\mathcal{F}_{\phi }^{th}=e^{-2(1+2n)\vartheta }.  \label{thermal}
\end{eqnarray}%
The dynamics of $\mathcal{F}_{\phi }^{th}$ with different temperatures are
plotted in Fig. \ref{temperature} with $\lambda =0.01\gamma $ based on
master equation (\ref{Eq-master}) in Panel (a) and based on Markovian
master equation given by Eq. (\ref{markovian}) in Panel (b). In both
conditions, one can find that the QFI with high temperature will decay more
rapidly than that with low temperature. Comparing Panels (a) with (b) for
the same temperature, one can also see that the QFI under Markovian limit
decays faster. Consider the relation between the QFI and the precision of
parameter estimation, i.e., $\delta (\phi )\geq 1/\sqrt{\nu \mathcal{F}%
_{\phi }}$, Fig. \ref{temperature} tells that 1) the precision of parameter
estimation will become lower if the reservoir gets hotter; 2) The Markovian
treatment will reduce the estimation precision.

\begin{figure}[tp]
\centering
\includegraphics[width=1\columnwidth]{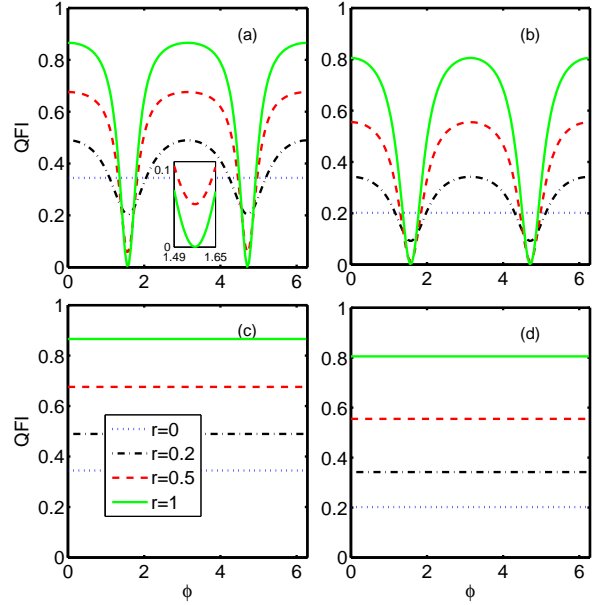}
\caption{(Color online) QFI vs. $\protect\phi$ at zero temperature. The QFI
in Panels (a) and (c) are given under the condition $\protect\lambda=0.1%
\protect\gamma $ and $\protect\gamma t=5$, while Panels (b) and (d) plot the
QFI under the Markovian limit at $\protect\gamma t=0.8$. The blue dotted line,
black dashed-dotted line, red dashed line and green line correspond to the
reservoir squeezing $r=0$, $0.2$, $0.5$, $1$, respectively. In (a) and (b), $%
\protect\theta=0$, and in (c) and (d), $\protect\phi-\frac{\protect\theta}{2}%
=0.01$ is satisfied for different $\protect\phi$ and $\protect\theta$ all
the time. The sub-Panel in (a) is the enlarged view in the vicinity of $\phi=\pi/2$.}
\label{matching}
\end{figure}

\subsection{The case with squeezing}

Comparing Eqs. (\ref{fisher}) and (\ref{thermal}), one can find that the
decay of QFI can be reduced, i.e., $\mathcal{F_{\phi }}>\mathcal{F}_{\phi
}^{th}$, if the following condition for $\phi $ and $\theta $ is satisfied,
\begin{eqnarray}
\cos ^{2}\left( \phi -\frac{\theta }{2}\right) >\frac{%
(A+B_{2}^{2}-1)(B_{1}^{2}-\mathcal{F}_{\phi }^{th})}{%
(B_{1}^{2}-B_{2}^{2})(A_{2}^{2}+\mathcal{F}_{\phi }^{th}-1)}.\label{ineq}
\end{eqnarray}%
This implies that the decay of QFI in the squeezed reservoir can be reduced
duo to the reservoir squeezing. This is obviously illustrated in Fig. \ref%
{matching}(a) and \ref{matching}(b) where the horizontal lines corresponds
to the case without squeezing. It is apparent that the squeezing makes a
large number of QFI surpass the horizontal lines, namely, the decay of QFI
has been reduced. In particular, one can see that with the increasing of
squeezing parameter $r$, the QFI is turning high and that the region below
the horizontal lines is getting small. In fact, this can be easily
understood in physics. In contrast to the case without squeezing, the
squeezing divided the imposed phase $\phi$ into two parts respectively related to
the squeezing relevant parameters $B_{1}$ and $B_{2}$ which has the opposite
behaviour with $r$ (see Eq. (\ref{canshu})). When $\phi -\frac{\theta }{2}\rightarrow 0$ or $\pi$, $B_{2}$
plays the dominant role in the quantum Fisher information due to the derivative
relation, which shows that the quantum Fisher information becomes large with the
increasing $r$. On the contrary, $B_{1}$ will play the dominant role. This
relation is obviously shown in Fig. \ref{matching} (a) and (b). That is, the  competition between the $B_1$ and $B_2$ lead to the reducing or increasing the decay of the QFI under reservoir squeezing for the different regions $\phi -\frac{\theta }{2}$. From the above  analysis, one can find that the squeezing phase parameter $\theta $ may play a significant role in the effects of the
precision of parameter estimation. We will show it in the following.

\subsection{Squeezing phase matching}

From the analytic QFI of parameter $\phi$ Eq. (\ref{fisher}), one can find
that when $\phi -\frac{\theta }{2}=0$ or $\pi $, the QFI can reach the
maximum with other parameters fixed. In this case, we say that $\theta$ and $%
\phi$ satisfy the \emph{squeezing phase matching}. If the estimated phase $%
\phi$ happens to be in the vicinity of squeezing phase matching, i.e., $\phi-%
\frac{\theta}{2}\leq\delta'$ with $\delta'$ a small quantity, one will find
that the decay of QFI will be prevented thoroughly, which can be found in
Fig. \ref{matching}(c) and \ref{matching}(d), where we assume $\delta'=0.01$.
The most obvious role of the (approximate) squeezing phase matching in Fig. %
\ref{matching} is that the regions below the horizontal lines in Fig. \ref%
{matching}(a) and Fig. \ref{matching}(b) are eliminated, which is just shown
in Fig. \ref{matching}(c) and \ref{matching}(d). In addition, one will find
that in Fig. \ref{matching}(c) and \ref{matching}(d), the QFI with squeezing
don't depend on the estimated phase $\phi$. The reason is that we have
chosen the same $\delta'$ for all $\phi$. In fact, it is not necessary to do
so. Thus we can conclude that the negative role of the reservoir squeezing
could be compensated by the squeezing phase matching.

\begin{figure}[h]
\centering
\includegraphics[width=1\columnwidth]{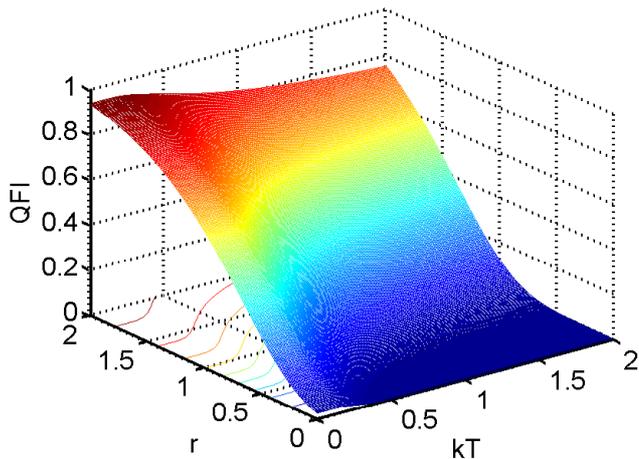}
\caption{(Color online) QFI vs the temperature $kT$ (in the unit of $\protect%
\omega$) and the squeezing parameter $r$ under condition $\protect\lambda=0.1%
\protect\gamma$ and $\protect\gamma t=10$. Here $\protect\phi-\frac{\protect%
\theta}{2}=0.01$ is satisfied for different $\protect\phi$ and $\protect%
\theta$ all the time. }
\label{simulaneous}
\end{figure}

In fact, Fig. \ref{matching} illustrates the different between the cases
vacuum reservoirs with and without squeezing, since we have set the
temperature to be zero. In order to find out the influence of temperature
accompanied by the squeezing, we plot the QFI in Fig. \ref{simulaneous} as a
function of the reservoir squeezing $r$ and the temperature $T$ at $\gamma
t=10$ under condition $\lambda =0.01\gamma$ and the approximate squeezing
phase matching condition $\delta'=0.01$. It is shown that with the squeezing
phase matching, the reservoir squeezing $r$ always plays the positive role
in restraining the decay of QFI, but the temperature plays a negative role.

\subsection{The effects of spectral property}

\begin{figure}[h]
\centering
\includegraphics[width=1\columnwidth]{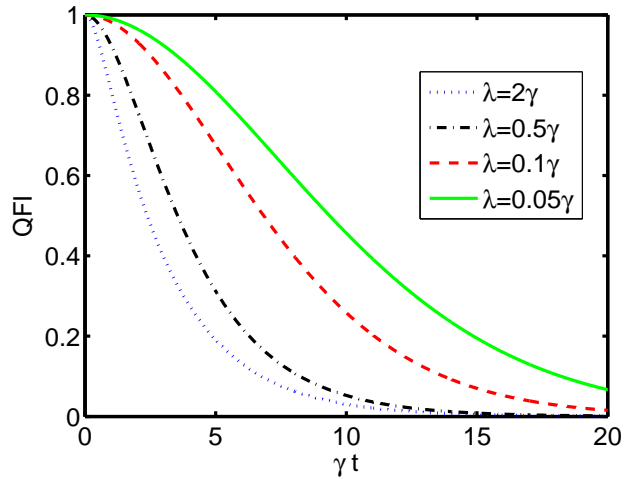}
\caption{(Color online) QFI vs $\protect\gamma t$ for various spectral
widths. Here we set $r=0.5$, $kT=0.5$ with squeezing phase
matching $\protect\phi-\frac{\protect\theta}{2}=0.01$ satisfied. The
spectral width $\protect\lambda$ are chosen as $\protect\lambda=2 \protect%
\gamma$, $0.5 \protect\gamma$, $0.1 \protect\gamma$, $0.05 \protect\gamma$,
respectively. }
\label{lambda}
\end{figure}

As is mentioned previously, the master equation (\ref{Eq-master}) does not
reveal the phenomenon of temporary information back flow \cite%
{hanshimaerkefu,feima-breuer,feima-breuser1,luxiaoming}. However, it does
not mean that the environment does not impact the dynamics of the QFI, which
can be easily found from Eq. (\ref{fisher1}). In order to intuitively
demonstrate such relations, we plot the QFI under different spectral widths $%
\lambda$ for $r=0.5$, $kT=0.5$ and squeezing matching condition $\phi-\frac{%
\theta}{2}=0.01$ in Fig. \ref{lambda}. One can easily find that the smaller
the spectral width is, the more slowly the QFI decays.

\section{Conclusion}

In summery, we have investigated the effects of reservoir squeezing on the
precision of parameter estimation based on non-perturbation procedures. The
exact analytic expression of the quantum Fisher information (QFI) is
obtained. The QFI depends on the estimated phase $\phi$ and the reservoir
squeezing parameter $r$, $\theta$. We have shown that the decay of the QFI
can be reduced by the reservoir squeezing, in particular, when taking into
account the \emph{squeezing phase matching}.

\section*{Acknowledgement}

We would like to thank Dr. H. Z. Shen and Jiong Cheng for fruitful
discussions. This work was supported by the National Natural Science
Foundation of China, under Grants No.11375036 and 11175033 and the Xinghai Scholar Cultivation
Plan.


\begin{thebibliography}{99}
\bibitem{metrology06prl} V. Giovannetti, S. Lloyd, and L. Maccone, Phys.
Rev. Lett. 96 (2006) 010401.

\bibitem{metrology11np} V. Giovannetti, S. Lloyd and L. Maccone, Nat. Phot.
5 (2011) 222.

\bibitem{fisher} H. Cram\'{e}r, Mathematical Methods of Statistics
(Princeton University, Princeton, NJ, 1946).

\bibitem{jiliangxuejiuchan} L. Pezz\'{e} and A. Smerzi, Phys. Rev. Lett. 102
(2009) 100401.

\bibitem{jiliangxuejiuchan2} P. Hyllus, et al, Phys. Rev. A 85 (2012) 022321.

\bibitem{wangxiaoguangzongshu} J. Ma, X, Wang, C.P. Sun, F. Nori, Phys. Rep.
509 (2011) 89.

\bibitem{xiaomin} L.J. Zhang and M. Xiao, Chin. Phys. B 22 (2013) 110310.

\bibitem{xiangguoyong} G.Y. Xiang and G.C Guo, Chin. Phys. B 22 (2013)
110601.

\bibitem{kexue} V. Giovannetti, S. Lloyd, and L. Maccone, Science 306 (2004)
1330.

\bibitem{PRA2012} K. Berrada, S.A. Khalek, and C.H. Raymond Ooi, Phys. Rev.
A 86 (2012) 033823.

\bibitem{CR-boud2} C.W. Helstrom, Quantum Detection and Estimation Theory
(Academic Press, New York, 1976).

\bibitem{Caves94prl} S.L. Braunstein and C.M. Caves, Phys. Rev. Lett. 72
(1994) 3439.

\bibitem{Caves} C.M. Caves, Phys. Rev. D 23 (1981) 1693.

\bibitem{winelind} J.J . Bollinger, W.M. Itano, D.J. Wineland and D.J.
Heinzen, Phys. Rev. A 54 (1996) 4649.

\bibitem{tigao-NOON} K.J. Resch et al, Phys. Rev. Lett. 98 (2007) 223601.

\bibitem{tigao-NOON2} J.A. Dunningham, K. Burnett, and S.M. Barnett, Phys.
Rev. Lett. 89 (2002) 150401.

\bibitem{tigao-coherence} J. Joo, W.J. Munro, and T.P. Spiller, Phys. Rev.
Lett. 107 (2011) 083601.

\bibitem{tigao-squeezed1} P.M. Anisimov, et al, Phys. Rev. Lett. 104 (2010)
103602.

\bibitem{tigao-squeezed2} L. Pezz\'{e} and A. Smerzi, Phys. Rev. Lett. 110
(2013) 163604.

\bibitem{xiangweipipei} J. Liu, X. Jing, and X. Wang, Phys. Rev. A 88 (2013)
042316.

\bibitem{Escher} B.M. Escher, R.L. de Matos Filho, and L. Davidovich, Nat.
Phys. 7 (2011) 406.

\bibitem{EscherPRL} B.M. Escher, L. Davidovich, N. Zagury, and R.L. de Matos
Filho, Phys. Rev. Lett. 109 (2012) 190404.

\bibitem{disange} R. Demkowicz-Dobrza\'{n}ski, J. Ko{\l }ody\'{n}ski and M.
Gu\c{t}\u{a}, Nature Commun. 3 (2012) 1063.

\bibitem{Plenio97} S.F. Huelga, et al, Phys. Rev. Lett. 79 (1997) 3865.

\bibitem{Plenio12} A.W. Chin, S.F. Huelga, and M.B. Plenio, Phys. Rev. Lett.
109 (2012) 233601.

\bibitem{majian} J. Ma, Y.X. Huang, X. Wang, and C.P. Sun, Phys. Rev. A 84
(2011) 022302.

\bibitem{sunzhe} Z. Sun, J. Ma, X.M. Lu,and X. Wang, Phys. Rev. A 82 (2010)
022306.

\bibitem{Fishe1} R. Krischek, et al, Phys. Rev. Lett. 107 (2011) 080504.

\bibitem{Fishe2} H. Strobel, et al, Science 345 (2014) 424.

\bibitem{Berrada} K. Berrada, Phys. Rev. A 88 (2013) 035806.

\bibitem{tigao-wangxiaoguang} Q.S. Tan, Y. Huang, X. Yin, L.M. Kuang, and X.
Wang, Phys. Rev. A 87 (2013) 032102.

\bibitem{tigao-redesign} L. Ostermann, H. Ritsch, and C. Genes, Phys. Rev.
Lett. 111 (2013) 123601.

\bibitem{tigao-errocode} W. D\"{u}r, M. Skotiniotis, F. Fr\"{o}wis, and B.
Kraus, Phys. Rev. Lett. 112 (2014) 080801.

\bibitem{tigao-nosie} R. Chaves, J.B. Brask, M. Markiewicz, J. Ko{\l }ody%
\'{n}ski, and A. Ac\'{\i}n, Phys. Rev. Lett. 111 (2013) 120401.


\bibitem{Gdadiner} C.W. Gardiner, Phys. Rev. Lett. 56 (1986) 1917.



\bibitem{finite-bandwidth} A.S. Parkins and C.W. Gardiner, Phys. Rev. A 37
(1988) 3867.

\bibitem{Zoller} A.S. Parkins, P. Zoller and H.J. Carmichael, Phys. Rev. A
48 (1993) 758.
\bibitem{fb} A.S. Parkins and C.W. Gardiner, Phys. Rev. A 40 (1989) 3796.
\bibitem{guanxisheng} M.M. Ali, P.W. Chen, and H.S. Goan, Phys. Rev. A 82
(2010) 022103.

\bibitem{huangxiaoli} X.L. Huang, T. Wang, and X.X. Yi, Phys. Rev. E 86
(2012) 051105.

\bibitem{geometricphas} A. Carollo, et al, Phys. Rev. Lett. 96 (2006) 150403.

\bibitem{yasuo-fourwave} R.E. Slusher, L.W. Hollberg, B. Yurke, J.C. Mertz,
and J.F. Valley, Phys. Rev. Lett. 55 (1985) 2409.

\bibitem{wulingan} L.A. Wu, H.J. Kimble, J.L. Hall, and H. Wu, Phys. Rev.
Lett. 57 (1986) 2520.

\bibitem{yasuo1} P. Tombesi and D. Vitali, Phys. Rev. A 50 (1994) 4253.

\bibitem{yasuo2} N. L\"{u}tkenhaus, J.I. Cirac, and P. Zoller, Phys. Rev. A
57 (1998) 548.
\bibitem{yasuo-canliangzhuanhuan} C. Vollmer, et al, Phys. Rev. Lett. 112
(2014) 073602.
\bibitem{yasuo2013} E. Shahmoon and G. Kurizki, Phys. Rev. A 87 (2013)
013841.

\bibitem{yasuo3} T. Werlang, R. Guzm\'{a}n, F.O. Prado, and C.J. Villas-B%
\^{o}as, Phys. Rev. A 78 (2008) 033820.

\bibitem{yasuo-yangyaping} W.J. Gu, G.X. Li and Y.P. Yang, Phys. Rev. A 88
(2013) 013835.

\bibitem{nature} K.W. Murch, et al, Nature 499 (2013) 62.


\bibitem{lujingjifen} A. Ishizaki, Y. Tanimura, Chem. Phys. 347 (2008) 185.

\bibitem{Breuer} H.P. Breuer and F. Petruccione, The Theory of Open Quantum
Systems (Oxford University Press, New York, 2002).

\bibitem{Scully} M. O. Scully and M. S. Zubairy, Quantum Optics (Cambridge
University Press, Cambridge, 1997)

\bibitem{wangfaqiang} F.Q. Wang, Z.M. Zhang, and R.S. Liang, Chinese Phys. B
18 (2009) 0597.

\bibitem{hanshimaerkefu} H.P. Breuer, Phys. Rev. A 70 (2004) 012106.

\bibitem{feima-breuer} H.P. Breuer, E.M. Laine,and J. Piilo, Phys.Rev.Lett.
103 (2009) 210401.

\bibitem{feima-breuser1} E.M. Laine, J. Piilo,and H.P. Breuer, Phys. Rev. A
81 (2010) 062115.

\bibitem{luxiaoming} X.M Lu, X. Wang, and C.P. Sun, Phys. Rev. A 82 (2010)
042103.

\end{thebibliography}
\end{document}